# Systematic Literature Review: Anti-Phishing Defences and Their Application to Before-the-click Phishing Email Detection

Trevor Wood[a*], Vitor Basto-Fernandes[b], Eerke Boiten[c], Iryna Yevseyeva[d]

[a]Faculty of Computing, Engineering and Media, De Montfort University, Leicester, UK, and director of Network Midlands Ltd, Leicester, UK

[b]Instituto Universitário de Lisboa (ISCTE-IUL), ISTAR-IUL, University Institute of Lisbon, Av. das Forças Armadas, Lisboa 1649-026, Portugal

[c]Professor of Cyber Security and Head of the School of Computer Science and Informatics, Faculty of Computing, Engineering and Media, De Montfort University, Leicester, UK

[d]Faculty of Computing, Engineering and Media, De Montfort University, Leicester, UK

*Abstract*—**Most research into anti-phishing defence assumes that the mal-actor is attempting to harvest end-users' personally identifiable information or login credentials and, hence, focuses on detecting phishing websites. The defences for this type of attack are usually activated after the end-user clicks on a link, at which point the link is checked. This is known as after-the-click detection. However, more sophisticated phishing attacks (such as spear-phishing and whaling) are rarely designed to get the end-user to visit a website. Instead, they attempt to get the end-user to perform some other action, for example, transferring money from their bank account to the mal-actors account. These attacks are rarer, and before-the-click defence has been investigated less than after-the-click defence. To better integrate and contextualize these studies in the overall anti-phishing research, this paper presents a systematic literature review of proposed anti-phishing defences. From a total of 6330 papers, 21 primary studies and 335 secondary studies were identified and examined. The current research was grouped into six primary categories, blocklist/allowlist, heuristics, content, visual, artificial intelligence/machine learning and proactive, with an additional category of "other" for detection techniques that do not fit into any of the primary categories. It then discusses the performance and suitability of using these techniques for detecting phishing emails before the end-user even reads the email. Finally, it suggests some promising areas for further research.**

*Keywords*—**anti-phishing defences, artificial intelligence, blocklist/allowlist, heuristic detection, machine learning, natural language processing, phishing, phishing detection techniques, phishing email detection, phishing website detection, proactive phishing detection, systematic literature review**

## 1. Introduction

A lot of research has been performed with the purpose of detecting phishing attacks. However, nearly all of this research is focused on detecting phishing websites that are being used to steal end-users' login credentials or pay for something they will never receive. (1) calls this "after-the-click" detection, that is, checking the URL or website after the user clicks on a link. Very little research has been performed with the aim of protecting end-users from advanced spear-phishing and whaling attacks. This systematic literature review starts by looking at anti-phishing defences that are currently being used in the real world and describing the lifecycle for combatting phishing attacks. It continues by determining what techniques are currently being used or proposed for use by automated defences to detect phishing attacks. It will discover that there are seven detection techniques. It will then examine these techniques to determine which ones could be used to examine emails before the user reads them or clicks on any links ("before-the-click" detection). In the real world, before-the-click detection is used widely to detect spam emails (2–4); however, spam detection cannot detect phishing emails as phishing emails are designed to look as much like genuine emails as possible. Finally, it will suggest several areas that could be researched further.

Please note that although we are using the terms blocklist and allowlist instead of blacklist and whitelist, we have not changed any quotes that use these terms.

### 1.1. What is phishing

Lexico defines phishing as "*The fraudulent practice of sending emails purporting to be from reputable companies in order to induce individuals to reveal personal information, such as passwords and credit card numbers. [as modifier] 'an email that is likely a phishing scam'*" (5). Other common definitions include "*… the process of attempting to acquire sensitive information such as usernames, passwords and credit card details by masquerading as a trustworthy entity using bulk email which tries to evade spam filters*" (6) and "*…a form of fraud in which an attacker masquerades as a reputable entity or person in email or other forms of communication*" (7).

In practice, phishing is an attempt by a mal-actor to induce a victim to act in a way that is beneficial to the mal-actor and detrimental to the victim. This could include persuading the victim to pay for something that seems legitimate but is not (the "Domain Renewal Group" scam (8)), to disclose bank details (Nigerian email scam (9)), or other personal information (username, password, etc.), downloading malware or opening files with malicious payloads (10). In recent years, the scams have been getting more sophisticated, using publicly available

* Corresponding author.
    Email addresses: trevor@the-woods.org.uk (T. Wood), vmbfs@iscte-iul.pt (V. Basto-Fernandes), eerke.boiten@dmu.ac.uk (E. Boiten), iryna@dmu.ac.uk (I. Yevseyeva)

information to tailor emails to look as though they are from the recipients' friends or family (Family Emergency Scams (11), Stranded Traveler Scam (12)).

Most recently, phishing has expanded from emails into voice phishing (vishing) and SMS phishing (smishing), as well as becoming more targeted through spear-phishing and whale phishing/whaling (13,14).

### 1.1.1. Spear phishing

"*Spear phishing is an email spoofing attack that targets a specific organization or individual, seeking unauthorized access to sensitive information*" (13). This attack is aimed at a small group of victims and appears to come from a known and trusted individual, for example, someone from within the victim's own organization or from a known and trusted supplier to the organization. Generally, the email also appears to come from someone in authority (the company director/CEO or the finance department/credit controller of the supplier). Very often, the mal-actor has researched information about the victim (for example, using social media, social engineering, etc.) and requires the victim to perform a manual action with a level of urgency (invoice needing payment, advance to cover expenses, etc.).

### 1.1.2. Whaling

Whaling (also known as Whale Phishing) is a phishing attack where the target victim is a high-level employee of a company (for example, the company director, finance director, CEO, etc.) (14). The aim is to manipulate the victim into divulging high-value information about the company or authorizing a high-value monetary transaction into the mal-actor's bank account. The attack often comprises of social engineering, email spoofing, and content spoofing and is hence exceedingly difficult for the intended victim to detect.

## 1.2. Why phishing is a problem

Cyber-attacks against businesses are on the increase worldwide. These attacks can be against all sizes of businesses, from a small local solicitor's firm leaving them unable to complete on clients' house sale and purchase transactions (15), through to a large multinational web hosting company which exposed the passwords of 1.2M users (16,17). The UK government published its latest Cyber Security Breaches survey in March 2021 (18). In their report, they state that 39% of UK businesses and 26% of UK charities reported having at least one cybersecurity breach or attack in 2020. Although there were fewer businesses reporting attacks (down from 46% in 2019), this is probably because of reduced trading due to the Covid-19 pandemic.

This report states that "*These figures have shifted gradually over time – the proportions experiencing negative outcomes or impacts in 2021 are significantly lower than in 2019 and preceding years. This is not due to breaches or attacks becoming less frequent, with no notable change in frequency this year*". The average cost of a cybersecurity breach is estimated to be £8,460 (up from £3,230 in 2019 (19)), although this is higher for medium and large firms (£13,400 up from £5,220 in 2019). Much of this money ends up in the hands of organized crime, both in the UK and overseas.

IT Governance reports that cybercrime has doubled in the last five years (20), costing UK businesses approximately £13 billion in 2019 and more than £87 billion in the last five years (21). They also state that almost half of all successful attacks against UK businesses started with some form of phishing attack. However, (22) reports that phishing accounts for 83% of attacks against businesses and 79% against charities. Either way, this makes phishing the most successful form of cyber-attack.

Note that the Department for Digital, Culture, Media and Sport survey does not include attacks against domestic end-users. In fact, it is difficult to find any figures for phishing attacks against domestic users. However, each year as the UK comes towards the tax self-assessment time, Her Majesty's Revenue and Customs (HMRC) issue a warning against tax-related scams. The latest warning stated that over 800,000 tax-related scams were reported over the last 12 months (23).

Similar trends are seen in the USA and Europe. In the USA, the FBI's Internet Crime Complaint Center (IC3) reports that the losses due to cybercrime in 2020 exceed $4.1 billion. $1.8 billion is due to Business Email Compromise (BEC) and $54 million due to phishing scams (24). Proofpoint report that 75% of organizations worldwide faced attempted phishing attacks, including spear-phishing, whaling, and BEC attacks (25). They then go on to state that 75% of US companies experienced a successful attack (30% higher than the global average).

In Europe, Allot reports that in 2020 phishing attacks increased by 718% compared to 2019 (26). However, it should be noted that Allot is a provider of security solutions, and this figure may be inflated due to improved detection techniques or marketing hype. ENISA (the European Union Agency for Cybersecurity) reports that the largest percentage of cybersecurity incidents were phishing (41%) (27). This figure is very similar to that reported by Kaspersky, which states that 40% of their users in the EU encountered at least one phishing attack (28).

## 1.3. Combatting phishing

There are currently two options for detecting phishing emails: software solutions and user detection.

Software detection is performed by the end-user's web browser, a browser plugin (if installed), or stand-alone software (if installed).

### 1.3.1. Browsers

(29) tests three of the most popular web browsers (Internet Explorer, Google Chrome, and Mozilla Firefox) to determine which one gives the best defence against phishing attacks. The browsers only perform one check, that is of comparing the URL with a blocklist. Although all the browsers provide some protection, none of them is foolproof as they only display a warning and still allow the end-user to proceed to the website.

Since (29) was published, support for Internet Explorer has been discontinued, and it has been replaced by Microsoft Edge. So far, it appears as though no one has evaluated the effectiveness of Microsoft Edge's anti-phishing defence, although it is likely to be very similar to that of Internet

Explorer.

### 1.3.2. Browser plugins

As part of their systemization of knowledge, (30) reviews nine different anti-phishing browser plugins (or toolbars). They do not attempt to evaluate the effectiveness of different browser plugins but just describe what protection they claim to offer. The mere fact that developers are prepared to create browser plugins and distribute them for free implies that the anti-phishing defences of browsers can be improved.

### 1.3.3. Software

Some stand-alone software is available to examine emails and help prevent phishing attacks. However, the commercially available software (for example, Mimecast[1], Proofpoint[2], Barracuda[3], Broadcom[4], Cisco[5], Fortinet[6], and Forcepoint[7]) is designed to be integrated into the mail systems of large companies. Small to Medium Enterprises (SMEs), micro-businesses, and domestic users cannot afford the cost of these systems. Some anti-virus/firewall software (for example, Bitdefender[8], Avast[9], Kaspersky[10], and McAfee[11]) includes the facility to check a URL against a blocklist, again after the end-user has clicked on the URL link. Like browsers and plugins, the end-user is warned that the URL may be malicious but is given the option to continue to visit the web page.

### 1.4. Reporting suspect emails

End-users can also report suspicious emails. Many large online companies (for example, eBay, PayPal, Amazon, Microsoft, Google Mail, banks, etc.) have a facility where end-users can report suspicious emails that appear to come from the company. In addition, there are two non-business organizations where end-users can report suspicious emails. In the UK, there is the National Cyber Security Centre's (NCSC) Suspicious Email Reporting Service (SERS)[12]. The NCSC is part of the UK's Government Communications Headquarters (GCHQ). Internationally there is the Anti Phishing Working Group (APWG)[13].

### 1.4.1. Commercial companies

Commercial companies are understandably unwilling to share their internal processes for processing suspicious emails that are reported to them. Most of them have information about spotting suspicious emails on their websites, and they usually send this off to the reporter. Some of these organizations collaborate with the NCSC and/or the APWG and forward the reported email to them for further examination.

### 1.4.2. NCSC and APWG

Suspicious emails sent to the NCSC and the APWG go through two or three further stages – analysis, information dissemination, and takedown.

#### 1.4.2.1. Analysis

In the analysis stage, suspicious emails are investigated to see if they are really phishing emails. The investigation by the NCSC seems to be limited to whether they contain links to phishing websites (new or already existing). NCSC does not perform the analysis themselves. Instead, they strip all Personally Identifiable Information (PII) from the email and forward it to their takedown provider, Netcraft[14], for analysis. The APWG assigns a "confidence" percentage to URLs indicating how certain they are that this is a phishing URL.

#### 1.4.2.2. Information Dissemination

If Netcraft ascertains that links in the suspicious email are pointing to phishing websites, that site is added to their database of phishing sites. This database is made available to partners who provide anti-phishing software, for example, anti-virus and firewall or browser plugin software developers, so the software can then scan for known malicious links.

The APWG add the link to their database immediately, along with the confidence percentage. They make this information available, in real-time, to their commercial and university research partners. They will also make available a dump of some of the data for researchers to evaluate before signing up as a partner.

The staff members at the NCSC that the first author interviewed said that only 10%-15% of emails reported are phishing emails (see Appendix A). The rest mostly comprises spam and junk emails.

#### 1.4.2.3. Takedown

The NCSC can request that a suspicious website be taken down by the hosting company. The UK government provides support for the removal of phishing websites that are hosted in the UK. If the site is attempting to spoof a government website (.GOV.UK), then this support is backed by law. UK hosting companies must comply with the request or face large fines. UK-based hosting companies are very cooperative when asked to remove any phishing website and often remove the site within minutes of it being reported. If the site is hosted overseas, the takedown is a little more complicated. The service provider is informed and asked to remove the website. In some countries, this request is ignored. However, access to the website can be blocked from the UK.

The APWG does not involve itself in takedowns, although some of their commercial partners attempt to do this, again with varying degrees of success.

### 1.5. Strengths and weaknesses of current defences

The NCSC system has its strengths and weaknesses. The primary strength is that it is backed by the UK government and can if required, use force of law within the UK. This backing

---

has encouraged UK-based hosting companies to comply with requests to remove phishing and spoofing websites.

There are two weaknesses. The first is variable co-operation from foreign governments, which ranges from good to completely ignoring any requests, especially by governments that are not friendly to the UK.

The second weakness is that the whole system relies on end-users spotting new or unknown phishing emails, which humans are particularly bad at doing.

The APWG system also has strengths and weaknesses. There are two strengths. The first is the number of high-profile commercial partners that work with the APWG to gather and disseminate information.

The second strength is that they also record the contents of phishing emails. The APWG will then make this data available to their partners and other cybersecurity researchers for further analysis.

The primary weakness is that the APWG has no power to take down phishing websites, nor does it attempt to.

Software defences also have their strengths and weaknesses. The strength is that they usually provide high true-positive and low false-positive results when detecting a phishing website. The main weakness is that software provides the option for the end-user to visit the site anyway. This is provided to overcome the false-positive results.

In addition, there are two other weaknesses to these systems: The first is that these systems only work for phishing emails that require the end-user to visit a website. There is no provision for spotting or protecting from advanced spear-phishing or whaling attacks where the end-user is asked to perform a manual task, for example, transferring money to a fraudster's bank account.

The second weakness is that anti-phishing software only blocks access to known phishing sites and hence provides no protection from zero-day attacks.

# 2. Literature review

## 2.1. Introduction

This section will examine what detection techniques are currently being used by automated defences. It will then group and examine these techniques with a view to applying them to "before-the-click" detection.

It should be noted that several of these techniques have only been described in academic research and have little or no proof of concept for use in the real world (30).

## 2.2. Methodology

This literature review is based on suggestions given by (31–33). The first section focused on reviewing current literature surveys and state-of-the-art publications in phishing detection. Most of the detection methods described and reviewed in these publications fall into one or more of the following techniques: Blocklist/allowlist, heuristic (or rule) based, email and website content based, visual comparison, artificial intelligence (AI)/machine learning (ML), proactive, and hybrid (a combination of two or more of the previous techniques). Two detection techniques do not fit into any of these categories, and therefore a category of "Other detection techniques" is used. The rest of the literature review is divided into these eight sections.

## 2.3. Phishing countermeasures

The following libraries were searched: De Montfort University[15], Scopus[16], IEEE[17], ACM[18], ProQuest[19], Wiley[20], Science Direct[21], and Google Scholar[22]. In addition, ResearchGate[23] was also searched to request the full text of two papers where only the abstract was available online. Web of Science[24] was excluded as it returned the same results as the De Montfort University Library.

The De Montfort University Library and Google Scholar were searched as a matter of course. The Scopus, IEEE, ACM, ProQuest, Wiley, and Science Direct libraries were included based on a Google search for publications that covered Cybersecurity issues.

The De Montfort University library included content from magazine articles, newsletters, newspaper articles, and trade publication articles. After examining the results returned in these categories, the categories were excluded from the search as they primarily provided real-life stories or advertorial material.

Note that the Google scholar advanced search is very basic compared to the advanced search facilities of the other libraries, in that it only allows searching within the title or the entire article. Searching of document abstracts and keywords is not supported.

Based on (31), the following steps were adopted to search for these sources. The details of each search are included in Appendix B.

### 2.3.1. Keyword search

The keywords for the initial search were phishing (in the title, abstract, or keywords) and "literature survey" OR "state of the art" OR SoK OR "Systematization of Knowledge" (in all available search fields).

### 2.3.2. Backward search

After the initial keyword search and screening of the papers identified in step 1, a backward search was performed looking for literature reviews referenced by those already identified. This search found several of the papers already identified but led to no new papers to investigate.

### 2.3.3. Forward search

Following the backward search, a forward search for citations of the publications identified in the initial search was

---

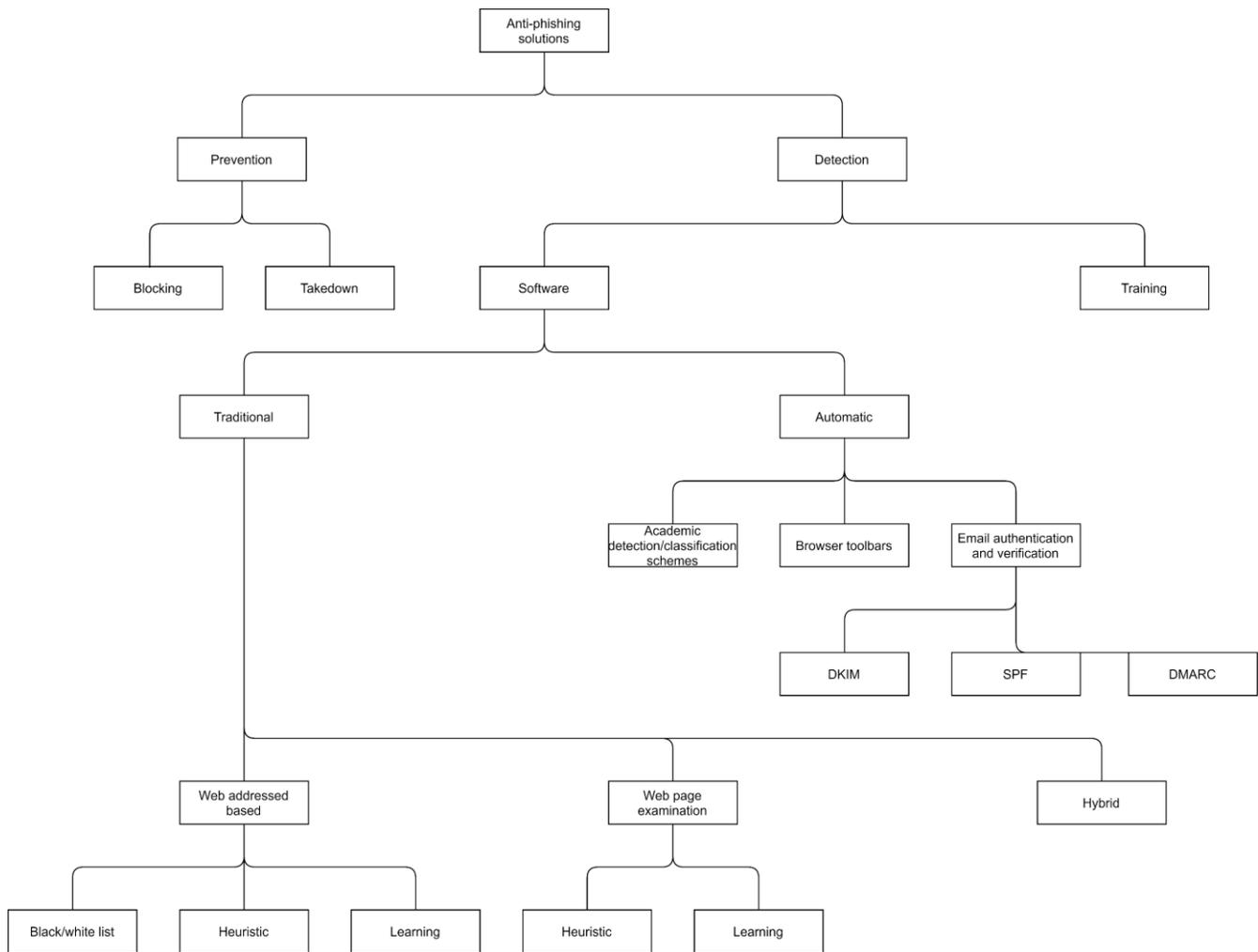

*Fig. 1 Phishing detection meta-hierarchy*

performed.

A further forward search of papers citing those found produced no new useful papers.

### 2.3.4. Keyword search for each of the identified detection techniques

Further searches of the libraries (excluding Scopus) were made for keywords associated with each of the identified detection techniques. The Scopus library search was dropped as all the search results were being returned from other libraries, and Scopus was only returning abstracts of the papers.

## 2.4. Grouping of phishing detection techniques

These publications examine many different methods for detecting phishing attacks. It must be noted that many of the authors restrict what phishing is in that they explicitly or implicitly define phishing attacks as attempting to entice the end-user to visit a phishing website. A fuller understanding could be "*Phishing is a type of computer attack that communicates socially engineered messages to humans via electronic communication channels in order to persuade them to perform certain actions for the attacker's benefit.*" (34). This definition would cover spear-phishing and whaling attacks – although it is noted that the example the authors give is still one enticing the end-user to visit a fake site. A non-website example, which the first author has seen, would be an accounts clerk receiving a final demand for an invoice from a supplier stating that if this invoice is not paid now, legal action would be taken. In this instance, the accounts clerk could not find the original invoice and so attempted to escalate the demand to the Financial Director of the organization. However, the Financial Director had started his vacation that day and was unavailable to confirm the demand. The accounts clerk took the demand to the IT manager, who, because of company policies, was not able to access the relevant accounts records. The result was the demand remained unpaid, and when the Financial Director returned from his vacation and checked the demand, it was determined to be a sophisticated spear-phishing attack that would have cost the organization £250,000 had it been successful.

It should also be noted that most of the literature written concentrates on detecting phishing websites after the user has clicked on a link and hence involves software (whether that be a web browser or separate stand-alone software such as a software firewall) intercepting the click communication, checking it, and blocking anything suspicious. This software may not be installed, or warnings could go unnoticed or be ignored by the end-user. In fact, (30) state that they "*do not cover phishing detection methods that perform email filtering because it is a different detection theme that warrants a separate comprehensive study on its own.*"

(35) classifies currently available phishing countermeasures into eight groups:

- stop phishing at the e-mail level
- security and password management toolbars
- restriction list
- visually differentiate the phishing sites
- two-factor and multi-channel authentication
- takedown, transaction anomaly detection, log files
- anti-phishing training
- legal solutions

Many of the countermeasures examined require human interaction (which is unreliable), the cooperation of the owners of legitimate websites to provide some form of authentication, or the intervention and cooperation of law enforcement agencies worldwide. However, several of the techniques examined in the "stop phishing at the e-mail level" and "restriction list" categories do not require any human intervention. Based on the interviews the first author had with members of staff at the NCSC and APWG, an examination of emails before the end-user reads them shows some promise in reducing the number of successful phishing attacks.

(36) describe a hierarchy of anti-phishing solutions, (37) propose a classification/taxonomy of anti-phishing techniques, and (38) propose a taxonomy for automated web phishing detection.

These hierarchies and taxonomies can be combined to describe a meta-hierarchy, as seen Fig. 1.

Other authors suggest different groupings of detection techniques. These include:

- heuristic-based, content-based, blocklist based, ML-based, and hybrid-based (39)
- search engine based, heuristic and ML-based, blocklist and allowlist based, visual similarity-based, Domain Name System (DNS) based, and Proactive URL detection based (40)
- basic features, latent topic model features, and dynamic Markov chain features (41)
- body-based features, subject-based features, URL based features, script-based features and sender-based features (42)
- blocklist, heuristics, visual similarity, and data mining (34)
- list-based (black/white), URL based (rules), content-based, hybrid, and image-based (43).
- blocklist, allowlist, heuristic (including rule-based classification, search engine assisted detection, content similarity detection, and visual similarity detection), ML, and honeypot (44).

Although this seems to be a large number of different techniques, these authors believe that they could all fit into one or more of these groups: blocklist/allowlist, heuristic, content, visual, ML, proactive, and hybrid. These groups will be used to examine and evaluate the various detection techniques discussed in these research papers. Additional research into these detection techniques will be discovered by extending the original literature review to focus on each detection technique in turn.

(30) propose a phishing detection technique based on Network Round Trip Time (NRTT). Although this method compares test results from two websites, it does not fit into any of the above groups, and so it will be discussed separately.

The concept of data mining as a technique was introduced by (45) and (46). The authors suggest that data mining could be used as part of heuristic, content, and ML groups of detection techniques.

Term Frequency/Inverse Document Frequency (TF-IDF) was identified in (30,47). TF-IDF is "*a statistical measure that evaluates how relevant a word is to a document in a collection of documents*" and is "*very useful for scoring words in machine learning algorithms for Natural Language Processing (NLP)*" (48). TF-IDF is used in the examination of website content.

# 3. Phishing detection techniques

This section examines the different detection techniques identified above in more depth, those techniques being grouped into blocklist/allowlist, heuristic, content, visual, AI/ML, proactive, other techniques, and hybrid.

## 3.1. Blocklist/allowlist

This section reviews the use of blocklists and allowlists of URLs, IP addresses, and DNS queries.

The keywords and results of this search are detailed in Appendix B.

A URL blocklist includes a list of URLs (either domain names or specific pages within a domain) that have been identified as malicious websites. These lists can include phishing websites, websites that distribute (or are infected with) malware, and badly behaved advertising sites. This technique has been used for a long time to protect end-users from many different threats and has been adopted by most after-the-click detection software to check the URL that has been clicked on. The URL that the end-user has clicked on is compared against the blocklist, and if the URL is found in that list, the end-user is warned about accessing that site.

There are numerous sources of these lists, including those compiled by PhishTank[25], PhishFindR[26], OpenPhish[27], the APWG, and the NCSC. This research does not evaluate the effectiveness of these lists, but much has been written comparing the different lists, for example (29,49–56).

URL allowlisting is an option that is used when a high degree of confidence is required, and it is rarely used. A allowlist contains a list of known good or legitimate URLs (either domain names or specific pages within a domain). In this instance, a URL is compared against the list and, if it appears in

---

| Blacklist | | |
|---|---|---|
| | Lack of zero-day (or zero-hour) detection | [53]–[70] |
| | Inability to detect the use of cloaking and evasion techniques | [71]–[73] |
| | Maintenance difficulties | [55], [64], [70], [74], [75] |
| | Inability to detect domain squatting (where phishing websites impersonate legitimate domains) | [59] |
| | Short life expectancy of phishing websites | [76] |
| **Whitelist** | | |
| | High false-positive results, where a legitimate site is not listed | [56], [58], [67] |
| | Can be tricked | [71] |
| | Cannot detect domain compromise where the URL or email address is unchanged but under control of a mal-actor | [77]–[80] |

the list, the end-user is allowed to access the website. Allowlisting works on the assumption that if the URL is not in the allowlist, it must be malicious.

(39) states that "*A blacklist includes a list of websites that are declared as spam… URLs present in the blacklist have denied access. This means a user cannot surf this webpage.*"

(40) states that "*The methods in this category utilize the whitelist of normal websites and the blacklist containing anomalous websites to detect phishing.*" Although they then go on to elaborate on how blocklists could be built, they do not explain how allowlists are created.

Two other types of blocklists, in addition to URL blocklisting, have been proposed.

In 2009 Nominum announced that it would be releasing "*a novel DNS security capability that functions like a spam blacklist, providing automated, real-time checking of DNS queries against a list of Web sites that are known to be malicious*" (57). In 2017 Akamail acquired Nominum (58). A search of their website shows no indication that they have continued with this service.

(40) discusses the use of DNS-based validation of the IP address of websites to determine whether they are phishing sites or not. This is done by comparing the IP address to a list of authentic (or allowlist) IP addresses, and (59) suggest using blocklists for IP addresses and sender email addresses.

### 3.1.1. Strengths

Both blocklists and allowlists are easy to implement and use (60).

Blocklists have a very low false-positive rate – that is, safe URLs being classified as phishing due to all the URLs on the list having been checked. Allowlists have a very low false-negative rate – that is, phishing URLs being classified as safe, again due to all the URLs on the list being checked.

### 3.1.2. Weaknesses

Both blocklists and allowlists suffer from major weaknesses, as detailed in Table I. Arguably, the biggest is their inability to detect zero-day attacks. Zero-day attacks are "*recently discovered security vulnerabilities that hackers can use to attack systems. The term "zero-day" refers to the fact that the vendor or developer has only just learned of the flaw –*

*which means they have "zero days" to fix it. A zero-day attack takes place when hackers exploit the flaw before developers have a chance to address it*" (61). In reference to a phishing attack, a zero-day attack is one that is being seen for the first time, and hence the phishing URLs have not yet been added to a blocklist.

Blocklists also suffer from the very short lifetime of phishing websites. Recent research has shown that a phishing website or page may only exist for a few hours or a day at the most before being moved to another web hosting company and domain (62). Although the content remains the same, the blocklisted URL and/or IP address change.

### 3.2. Heuristic

This section reviews the use of heuristics (or rules) in examining websites and emails to detect phishing websites and emails.

The keywords and results of this search are detailed in Appendix B.

This technique uses heuristics (or rules) to detect potential phishing attacks. Most research has concentrated on after-the-click detection, where heuristics is used to classify the link URL or website characteristics. Essentially the URL is analysed through a set of rules, with each rule being used to determine the probability of the URL being phishy. However, the same rules could be applied to before-the-click detection by examining URLs in an email.

(44) suggests that the heuristic approach can be divided into four categories: rule-based, search-engine assisted, content similarity, and visual similarity. As there is quite a lot written about content and visual similarity, these techniques will be dealt with separately.

Several authors combine the heuristic detection technique with other detection techniques such as blocklists and/or allowlists (63–66), search engine ranking (67–69), web page examination (70), brand name impersonation (64), and ML/AI (66,71–78) to improve the percentage of phishing attacks detected. In most cases, rules are applied to extract features from the URL, email content, or web page content which is then used as input into the other techniques to ascertain a phishyness score.

| Category | Detection technique | Comments |
|---|---|---|
| Lexical properties | Long URL | |
| | Multiple sub-domains | |
| | Request URL | URL shorteners and mass email marketing software complicate this |
| | URL of the link anchor | URL shorteners and mass email marketing software complicate this |
| | Server form handler | Problematic with increased use of cloud services |
| | Pop-up windows | Should not contain a form that collects sensitive information. Complicated by cookie choice pop-up windows |
| | PHP and JavaScript scrips | mail() function or mailto: to submit user information |
| | Non-standard port as part of the URL | |
| | Abnormal prefix and/or suffix | |
| Obfuscation | URL encoding. Using hexadecimal character codes or Unicode (homographic attack) | Replacement of ASCII characters with URL encoded form is sometimes required |
| | Use of IP address | Replacing the domain name with the IP address |
| | URL shortening | Use of publicly available or custom URL shorteners [94], [95] |
| | Use of special symbols in URL ('-' dash, '_' underscore,, ',' comma, ';' semicolon, '@' at) | '-' (dash) and '_' (underscore) are legitimate symbols in a URL. It depends upon what surrounds it, e.g., pay-pal. |
| | URL redirection. Use of // anywhere other than after the URL protocol. Use of HTTP status codes 3xx | Complicated with the use of URL shorteners and mass email marketing software. Need to follow redirection to the end. Web services are available |
| Hostname and DNS properties | DNS record not available | |
| | Country-code validation. Is the country code different from the country where the site is hosted? | The proliferation of generic TLDs not assigned to any specific country |
| URL host features | WHOIS server details | Domain name registration, update, expiration dates, or no record available |
| | Use of HTTP/HTTPS | SSL certificates are available for free [96]. Over 80% of phishing sites now use them [97] |
| | Amount of website traffic as found by Alexa | Traffic can be faked |
| | Geographic information | |
| | Domain name similarity | |
| | Indexed in Google | Unfound by Google is a possible indicator |
| | Page/domain rank (Google) | Can be fooled with compromised domain |
| | Domain popularity (Alexa) | Can be fooled with compromised domain |
| | Number of inbound links | Can be faked |

(40) links heuristics with ML and states that this technique "*extract(s) a set of features of either text, image, or URL-specific information from normal or abnormal websites. A set of heuristics is utilized, and the thresholds or rules obtained from the learning algorithms are used for anomaly detection.*". They imply that heuristics can only be used with other techniques to become a hybrid technique, which is clearly incorrect, as can be seen from the references above.

Many of the sources suggest URL features that could be examined (30,39,69,79–103). Table II summarizes the different URL detection techniques within each category and highlights some potential issues with these techniques.

This technique could be extended to examine URLs and all addresses in the header and body of an email for characteristics that might prove to be phishy.

In addition to checking URLs in an email, several characteristics of the email header and body could be examined for potential phishing indicators (59,79,99,101). Header checks could include sender-related fields (From, Sender, Mail From, Reply-to, etc.), subject, received field, and others. Body checks could include keyphrases (for example, "Click here", "Urgent", "Warning", "Account closure notice", etc.), semantic features (for example, language that carries a sense of urgency or warning), and others. Semantic features require the use of

| Email header checks |
|---|
| "Message-Id" field |
| "Received" field |
| "From" field |
| "Mail from" field |
| "Sender" field |
| "Mail-to" field |
| "Delivered To" field |
| Authentication-Results (SPF, DKIM, etc.) |
| "Subject" features (length of subject, number of words, number of characters, vocabulary richness, etc.) |
| Blacklisted words in "Subject" |
| number of words and/or characters in the "Send" field |
| "Sender" domain ≠ "Reply-to" domain |
| "Sender" domain ≠ "Message-Id" domain |
| "Sender" / "From" ≠ email's modal domain |
| Timestamp, "Sent date" |
| Source IP, Autonomous System Number |
| "Subject": Fwd, Reply |
| Interaction habits |
| "Cc", "BCc" fields |
| "X-Mailer" field |
| "X-Originating-IP" field |
| "X-Originating-hostname" field |
| "X-spam-flag" field |
| "X-virus-scanned" field |
| Domain health: Issues with DNS, DMARC, SPF, DKIM, etc. |

| Email body checks |
|---|
| Lexical Features (number of words, number of characters, Function words, Tokens, Regular expressions, etc.) |
| Style metrics (number of paragraphs in the email, Yule metric, etc.) |
| Topic in the body |
| Latent Semantic Indexing |
| Readability indexes |
| NLP (Part-of-Speech tags, Named entities, Wordnet properties, etc.) |
| Semantic Network Analysis |
| Vocabulary Richness |
| Presence of urgency, reward, threat language, blacklisted words in the body |
| Greeting, signature, farewell in the message |
| Presence of both "From:" & "To:" in the email body |
| number of links to the domain |
| URL features |
| HTML features |
| Presence and/or number of forms in the email body |
| Blacklisted words in the message |
| Script/JavaScript features in the email body |
| number of onClick events in the email body |
| Feature from images/logos in the message |
| Mention of the sender |
| number of links & images links in the email body |
| number of tables in the email body |
| Recipient's email address in the email body |
| Phishing terms weight |
| TF-IDF |
| Email size |
| number of email body parts |
| MIME Version or Content-Type |
| JavaScript Popup Windows |
| Link displayed ≠ Link in destination |
| IMG links ≠ spoofed target address |
| Hidden text in the email, Salting techniques |
| number of internal/external links |
| Images used to look like file attachments and point to a URL link |

Natural Language Processing (NLP) techniques and hence will be discussed later. Other features may require the use of ML algorithms for detection and hence will also be discussed later. Roger Grimes, a Data-Driven Security Evangelist at KnowBe4, produced a webinar called "Cyber CSI: Learn How to Forensically Examine Phishing Emails to Better Protect Your Organization Today" (104). In this webinar, he described several techniques to forensically examine a suspicious email to determine whether it is phishy. Although these techniques are performed manually, some of them could be automated. Most of the techniques have already been identified, but he does suggest some additional checks that could be performed. Most of these checks are only viable if the email is claiming to come from a "big name" company, for example, Microsoft, IBM, Maersk, etc., and would probably be difficult to implement in a generalized email software checking service.

Table III and Table IV summarize the heuristic checks that could be performed on the email header and email body

In after-the-click detection, the web page is also examined (80,86,87,96,103,105–118) and Table V summarizes the heuristic checks that could be performed on a web page. Although these checks are specifically for web pages, it may be possible to use them in checking emails or as part of proactive detection techniques.

Two papers examine the process for comparing a potentially phishing URL with the "real" URL – for example, is login-paypal.com really part of paypal.com or is it a phishing page

(119,120). However, their techniques require knowing what the real site is. This may be straightforward for well-known and often imitated brands like PayPal and Amazon, but the authors do not explain how this could be used when the real site cannot be ascertained.

A special mention needs to be made on homographic attacks. In a homographic attack, one or more characters (or glyphs) are replaced with ones that look similar but from another language set. This is made possible with the use of Unicode. Unicode is an "*international character-encoding system designed to support the electronic interchange, processing, and display of the written texts of the diverse languages of the modern and classical world*" (121). Because Unicode covers the symbols in almost all world languages, it is quite possible to find symbols that look visually similar, as seen

| Web page feature | Comment |
|---|---|
| Tracking of a login screen | Does it attempt to get any login information from the user |
| Status/address bar customization | It is possible to display a fake URL in the status/address bar |
| MouseOver | on_mouseover event used |
| Disabling right-click | Disables the viewing of source code |
| Pop up | Not used to collect login information. Note this is not necessarily true. Some legitimate sites pop up a login window |
| Iframe redirection | "hiding" a phishing page inside another page to make it invisible |
| Links in <meta>, <script> and <link> tags | Links to another domain. However, the use of Google Analytics, Google reCAPTCHA, FontAwesome, etc. would break the rule |
| Favicon mismatch | Favicon image referenced from another domain |
| Cookies | Cookies are not used on phishing websites. As stated by [65] but no evidence has been supplied |
| Logos and images | Are they hosted by the same domain? If not, why not? |
| Hidden/restricted information | Does the page use <div style = "visibility: hidden"> or <div style="display: none">? Sometimes used for drop-down or accordion style menus. <input type = "hidden"> hides the input box, <input disabled = "disabled">. Sometimes used to pass additional information (such as page URL) for processing |

| Real "apple.com" | | | Fake "apple.com" | | |
|---|---|---|---|---|---|
| Glyph | Unicode name | Unicode hex | Glyph | Unicode name | Unicode hex |
| a | Latin lowercase A | U+0061 | a | Cyrillic lowercase A | U+0430 |
| p | Latin lowercase P | U+0070 | p | Cyrillic lowercase Er | U+0440 |
| l | Latin lowercase L | U+006C | l | Cyrillic lowercase Palochka | U+04CF |
| e | Latin lowercase E | U+0065 | e | Cyrillic lowercase Ie | U+0435 |

at Unicode Utilities: Confusables[28]

These visually similar characters can be used to fool the end-user. For example, a false apple.com could be set up by replacing the Latin lowercase letters A, P, L, and E with Cyrillic lowercase letters A, Er, Palochka, and le. To the human eye, these letters (or glyphs) look visually similar (see Table VI) but would take the user to a completely different web page.

Very recently, a modified version of the homographic attack has been discovered where the mal-actors use mathematical symbols to impersonate a company's logo (122).

Although several authors have identified homographic attacks as a problem (102,123,124), only two papers suggest potential solutions (125,126), and neither seems to have demonstrated their defence in anything other than a simulation.

Finally, in this section, shortened URLs need to be discussed. In 2002, the first URL shortener, TinyURL[29], was launched. Its raison d'être was purely to shorten long and complicated URLs to make them more usable. Over the following years, many other URL shorteners have been released – some by social media and search engine giants using part of their name as the URL (for example, Google uses goo.gl, YouTube has youtu.be and LinkedIn has lnkd.in) and some to make use of limited space (for example Twitter's famous 140-character limit). Some of these URL shorteners are reported to be actively used in attacks (127). Nowadays, anyone can set up their own URL shortener (128,129), and mal-actors have done this specifically to avoid some of the heuristic rules above. Although some websites can follow redirects from URL shorteners (for example, Redirect Detective, Redirect Checker, and WhereGoes[30]), they require a URL to be manually inputted on their websites. There are also browser plugins that perform the same task, but they do need to be installed by the end-user. There is also no indication that these websites or plugins can check custom build URL shorteners.

### 3.2.1. Strengths

Heuristics (or rules) provide responses that can be accurately measured. However, some rules may provide a stronger or weaker indication that the item being tested (URL, email, website) is phishy, and humans need to pre-assign a score to the result.

### 3.2.2. Weaknesses

Mal-actors could find ways to defeat this technique, either by crafting the attack to pass the rules or bypassing the rule checking by creating an attack that does not fit any of the rules.

### 3.3. Content

This section reviews how the examination of the content of websites, web pages, and emails can be used to detect phishing

---

[28] https://util.unicode.org/UnicodeJsps/confusables.jsp
[29] https://tinyurl.com/app

[30] https://redirectdetective.com/, https://www.redirect-checker.org/, https://wheregoes.com/

websites and emails using methods other than rules.

The keywords and results of this search are detailed in Appendix B.

In after-the-click detection, the most common content checking technique used is to compare two web pages, the original known good web page and the suspected phishy one. Some content checking techniques use a visual approach and will be discussed in the next section.

(39) states that "*The comparison of two web pages is done based on the similar contents on the web page. This technique makes use of Term Frequency/Inverse Document Frequency (TF-IDF). TF-IDF compares the terms in the original website to the phishy one*". (130–132) describe how using TF-IDF is used in a content-based approach to detecting phishing websites called CANTINA. As CANTINA uses several other techniques, it could arguably be classified as a hybrid technique. (133) later extend this into CANTINA+ by using other detection techniques, moving it squarely into the hybrid category.

(40) introduces the idea of using search engines to extend content testing. They state that this technique "*extract(s) features such as text, images, and URLs from websites, then search for them using single or multiple search engines and collect the findings.*" Their assumption is that "*when detecting a normal website... it will be among the top search results, as normal websites typically have a higher index than phishing webpages, which remain active for a very short time*". However, there is evidence to show that search engines, especially Google, are showing search results for, what the BBC calls "shyster" websites, that is, websites that are charging for free services (134,135). (136) extends this idea with their tool GoldPhish. This tool captures an image of the web page and converts it to text using Optical Character Recognition (OCR) software. The extracted text is then submitted to the Google search engine, which returns several URLs. The first four are then compared with the URL of the submitted page, and if there is no match found, the page is flagged as phishing. This seems like a very slow and clumsy method, as the text could be extracted automatically from the page source.

A slightly different approach, but again using Google search, is proposed by (137). Again, they suggest taking a screenshot of the website in question, but this time, instead of OCRing the text, the website logo is extracted and input into a Google Images search. Again, the resulting returned URLs are compared to that of the website, and if no match is found, the website is marked as phishy. The big, obvious problem with this approach is how to locate the logo in the screenshot.

Another variation is proposed by (138). Their solution is to extract brand names from the suspicious website and submit those found to Google. Again, the resulting returned URLs are compared with the suspicious website URL. However, to implement this, a list of brand names that could be the focus of an attack would have to be kept. Some research would have to be performed to ascertain the most attacked brand names.

Many websites today have a Favicon. This icon is shown in the browser tab when a page from the website is open. It is also stored with the bookmark when the page is added to the browser bookmarks. It assists the end-user in identifying a site quickly.

Favicons are images that are stored on the webserver with the website and hence have (or should have) the same domain name as the web page. (139) suggest that if the location of the favicon is different from the location of the page being inspected, then that page is phishy. This may work; however, mal-actors could (and probably do) circumvent this indicator by placing a copy of the favicon from the site they are trying to spoof within their own website.

Several authors suggest examining the web page's Document Object Model (DOM) (115,140,141). In this case, the DOM of a suspicious page is extracted, and if the DOM is found to be the same as a previous phishing page, the page being tested is flagged as phishing.

(142,143) both suggest using MD5 hashing to identify phishing websites. MD5 is a one-way cryptographic function. The idea behind it is that any string of characters can be put through the hashing algorithm and produce a unique 128-bit hash or fingerprint (144). (142) suggests that the source of known phishing web pages could be put through the MD5 algorithm, and the resulting hash stored. Later, the source of a suspicious web page could also be hashed and compared to the stored hashes, and, if it were a match, this would show that this web page is also a phishing page. Recently, the MD5 hash algorithm has been shown to not produce unique hashes (145,146). However, the idea has merit for detecting both phishing web pages and emails and should be explored further using the latest cryptographic hash algorithm, currently SHA-3 (147).

Imperceptible watermarking of text and images is proposed by (148). This method requires a URL-dependent watermark to be embedded in the elements of a web page so that when the web page is retrieved, the URL of the web page can be used to recreate the watermark, and any difference would indicate a phishing web page. The problem with this is that it requires all website developers to embed the watermark, and mal-actors could just do the same. There is also an issue connected with dynamic web pages – that is, web pages that are created using a server-side or client-side script. The authors claim that their proposal works for dynamic pages but do not explain how or offer any proof.

(149) suggests an approach using document authorship techniques. They propose using a system they call Anti-Spear phishing Content-based Authorship Identification (ASCAI). ASCAI detects possible mismatches of writing styles in emails purportedly sent from the same person, the idea being that a particular sender is likely to write in the same style, use the same sentence structure, and use the same punctuation styles when they write (for example a person may never use a semi-colon, or they always use the Oxford comma). When the body of an email is examined, the receiver is warned if there are significant differences. While this method would be difficult (if not impossible) to use for everyday phishing emails, it could be used within a large company where a lot of internal emails are being processed to help prevent spear-phishing attacks.

### 3.3.1. Strengths

This could be a good technique for detecting phishing

| Detection method | References |
|---|---|
| Page rank using image search | [36], [161] |
| Comparison with known phishing websites | [177] |
| Style comparison | [178], [179] |
| Visual similarity | [161], [174], [179], [181]–[199] |

attacks where the genuine website has been cloned.

### 3.3.2. Weaknesses

Many of the techniques described above will only work on examining web pages, which begs the question, "How is the genuine website determined for comparison purposes?"

## 3.4. Visual

This section reviews how automated visual examination of websites can be used to detect phishing websites.

In after-the-click detection, there are some similarities between content and visual comparison checking techniques, and several authors include the two together (150–152). However, there are enough differences to justify treating them as separate techniques.

The keywords and results of this search are detailed in Appendix B.

(40) states that this technique "*utilizes the visual similarity between webpages to detect phishing. When phishing web sites are matched in terms of their visual characteristics with the authentic websites, it checks whether the URL is on the authentic domain URL list. If not, the website is marked as a phishing website*".

Most of the research connected with this detection technique focuses on comparing the visual representation of the suspicious website or web page with the legitimate one. There appear to be four main methods for performing this comparison.

The first method is described by (39) as "*capturing the screenshots of a website and then process them to compare. This information retrieved from screenshots after processing can be given to a search engine to acquire its page rank and check the legitimacy of the website by comparing the content on it. Website logo can be used in this method to analyze the webpage using the Google image database*". Their summary is based on a detection method described in (137). In summary, this detection method assumes that the mal-actors will attempt to clone a real website in order to deceive the victim. They propose identifying the site logo on the suspect page and performing a Google image search to locate the highest-ranking page with that logo on it; they assume that the highest-ranking page will be the legitimate website. If the domain name of the website found in this search is not the same as that originally supplied, then the original target website is assumed to be phishy.

The second method is described in (153). In this method, the assumption again is that mal-actors will attempt to clone or mimic a legitimate website. This time, however, the comparison is made against other known phishing websites in an attempt to find a match. If a match is found, then the suspicious website is marked as phishing.

The third method is to compare the styles used in the suspicious website against those of the legitimate one. Again the assumption is that the sites must look visually similar in whichever web browser the intended victim uses. (154) states, "*To maintain a consistent look across browsers, attackers also need to rely on the CSS* [Cascading Style Sheet] *technology, as the target website does. Attackers should use CSS that results in visual appearance similar to that of the target page to lure users successfully.*" Their analysis showed that most phishing websites (as supplied by PhishTank[31]) either link directly to the CSS of the legitimate website or copy the CSS to the phishing website. Their method "*will extract the key features in CSS and page contents, and use them as a basis to decide the similarities of pages.*" (155) extend this detection method to examining blocks of text and images and layout similarities.

The fourth and most common method is to compare the visual similarity between the suspicious page and the legitimate page. Several ways of doing this are suggested and are described below. These are summarized in Table VII.

Most authors suggest comparing images found on the suspicious page to the images found on the legitimate website.

This is like the first method but assumes that the legitimate website can be determined, although none of the authors suggest how this could be achieved. Two other options have been suggested for website comparison, and these are:

Earth Mover's Distance (EMD): EMD is "*A semantic measure for document similarity in semantic search*" (156).

Arguably, its most common use is with search engines as they interpret a user's search to return the most relevant results. Although its main use is with words, there is a growing use of EMD for image retrieval (157) and (158) propose using EMD to calculate the visual similarity of web pages, the suspicious one and the legitimate one.

Gestalt theory: Gestalt Theory (or Psychology) suggests that the human mind does not look at items individually but sees them as part of a greater whole (159). (160) propose using this theory for detecting phishing websites by examining the web page as a whole and not breaking it down into different elements (for example, images, text, styles, etc.).

The visual detection methods are summarized in Table VII

Several hybrid approaches are proposed, where different detection techniques are used in an attempt to improve the speed and accuracy of detection, including pre-filtering using blocklists/allowlists (161,162), web page content examination (150–152), and heuristics (152,162,163).

In addition, (164,165) propose extracting visual elements as the first stage of the ML technique, which will be discussed later.

---

Although most of these authors claim that their visual detection methods produce a high detection rate with a low false-positive rate, their testing methods call into question the reliability of these results. The authors select known phishing websites and then compare them with the site that they are supposed to be imitating. In the real world, this would be unlikely to happen as the site being tested probably will not be a phishing site, and even if it is, there is then the issue of determining which legitimate site it is trying to impersonate. The only method that is likely to produce reliable results is the page ranking using image search.

### 3.4.1. Strengths

This technique could be useful to assist in the proactive detection of phishing websites where time and computing power are not so much of an issue and where human intervention could be requested to provide information about the legitimate site that is being imitated.

### 3.4.2. Weaknesses

There are three major weaknesses with the visual comparison technique. The first two are highlighted by (166) in that image comparison takes more time for comparison and more space for the storage of images than other detection techniques. (167,168) attempt to address the issue of speed by using two detection phases. They correctly assume that most pages a user visits are benign, so this process looks first to rule these out before passing any suspicious pages on to the next step, in this case, around visual inspection techniques.

The third, and probably the most major, weakness is not mentioned by any of the authors, that of determining what the legitimate website is to compare against. Nowhere do any of the research papers explain how they determine which website page or brand is being spoofed or how to find the legitimate website even if this information is known. All the research appears to be done assuming this knowledge is already known.

For this to be a viable technique for even after-the-click detection, somehow the webpage or brand being spoofed must be extracted from the suspect page, and then this must be linked to the legitimate website, presumably through some form of lookup table or Google search.

Because of these weaknesses, the authors suggest that this technique is not viable in the real world for after-the-click detection, let alone for examining emails in before-the-click detection.

## 3.5. Artificial intelligence/machine learning

This section reviews the use of AI and ML in examining websites and emails to detect phishing websites and emails. It includes deep learning and NLP as part of AI/ML. Some of the keywords used in the search were suggested from papers examined in earlier sections.

The keywords and results of this search are detailed in Appendix B.

### 3.5.1. Machine learning

ML is a branch of AI. Computer scientist and ML pioneer Tom M. Mitchell[32] defines it as "*...the study of computer algorithms that allow computer programs to automatically improve through experience*" (169). These algorithms learn by experience, similarly to the way humans learn. For example, an algorithm could be shown many pictures of dogs, told that they are dogs, and it will learn to recognize a dog in a picture it has never seen before (170). There have been plenty of introductions to ML written, for example (171–175).

In this technique, the software is taught how to detect phishing websites or emails.

ML can be implemented using three different learning processes: Deductive, Inductive, and Explanation-based learning (176).

According to (101), most ML algorithms can be grouped into three categories: Supervised, Unsupervised, and Online learners. They also indicate that some research has been done using rarely-used methods, for example, associative rule mining and Markov models. Interestingly, they also include rule-based (or heuristic) as a learning approach. From their literature review, it is clear that a majority of research is being focused on supervised learning. They conclude that other learning approaches are largely unexplored.

(39) states, "*In this technique* [supervised learning], *features are extracted and they are classified using the machine learning techniques.*"

(177) describes nine different ML techniques: Naïve-Bayes, Random Forest, Decision Tree, Support Vector Machine (SVM), K-Nearest Neighbor (KNN), Logistic Regression, Adaptive Boosting (AdaBoost), Artificial Neural Networks (ANN), and Fuzzy Logic. They further sub-classify ANN into Deep Neural Network, Feed-Forward Neural Network, Convolutional Neural Networks (CNN), Radial Basis Function (RBF), Multi-layer Perception (MLP), Recurrent Neural Networks (RNN), Long short-term memory (LSTM), Capsule Neural Network (CapsNet), Deep Belief Network (DBN) and Auto-encoder (AE). It is beyond the scope of this paper to explain the different ML classifiers.

Although not overtly stated in any of the research papers, it would appear that inductive learning is the primary learning process used in ML (178). In this learning process, data is supplied along with an indication of whether it is phishing or not. The algorithm is told what features to look at and then creates rules that can be used to determine whether unknown input is phishing or non-phishing. In effect, the ML algorithm mimics the human process of creating rules as described in the section on heuristics. Unfortunately, none of the authors reveals what rules their ML training have created, and so it is difficult or impossible to decide whether ML is creating rules that are comparable to those created by humans.

Several papers have been written where the results of different ML algorithms have been compared for accuracy. Unfortunately, different authors have compared different algorithms, used different datasets, and used slightly different ways of calculating the accuracy of the algorithms, making it difficult to compare across research. However, an examination

---

of (74,164,179–200) clearly shows that Random Forest and ANN perform consistently highly while Naïve Bayes, AdaBoost, and Logistic regression perform poorest.

(188,196,198,200) also show that using different classifiers affects the accuracy of the detection, although, in most cases, by less than one percent.

(183,188) demonstrate that the detection accuracy can be increased by combining multiple ML algorithms, albeit by less than 0.75%.

There is also evidence to suggest that data mining ("*the process of posing queries and extracting patterns, often previously unknown from large quantities of data using pattern matching or other reasoning techniques*" (201)) could be combined with ML to assist in the detection of phishing websites (202,203). However, the conclusions drawn by (202) indicate that, at the time of writing (2016), "*the methods that are the most effective for cyber applications have not been established; and given the richness and complexity of the methods, it is impossible to make one recommendation for each method, based on the type of attack the system is supposed to detect.*"

### 3.5.2. Deep learning

A further subset of ML is Deep Learning. Unlike ML, which requires the training data to be tagged as phishing or non-phishing, deep learning uses ANN to simulate how the human brain works and hence improve the way computers can learn (204–212).

### 3.5.3. Natural language processing

NLP is possibly the most interesting and least researched method of detecting phishing emails. NLP is another subset of AI which gives a computer the ability to understand, manipulate and respond to human language, whether written or spoken (213–217). It is most often portrayed in computers, robots, and androids in Science Fiction, for example, Ash in Alien (218), the replicants in Blade Runner (219), Huey, Dewey, and Louie in Silent Running (220), R2D2, C3PO, and other robots in the Star Wars franchise (221), the various terminators in The Terminator franchise (222), Master Control in Tron and Tron: Legacy (223,224), the talking, philosophical bomb in Dark Star (225) and possibly most famously The Gunslinger in Westworld (226), Data and the Enterprise's computer in Star Trek: TNG (227) and HAL in 2001: A Space Odyssey (228).

In the real world, NLP is being used for spam email filtering, smart assistants (like Siri and Alexa), search engine results, predictive text (autocomplete and autocorrect), language translation, digital phone calls ("This call may be recorded…"), data analysis and text analysis (229), with other applications being developed regularly.

Salloum et al. have compiled a fairly comprehensive literature survey of research into how NLP could be used to detect phishing emails (230). Most of the research has focused on low-level phishing attacks; however, (231) examines how NLP can be applied to the more complicated task of detecting spear-phishing attacks.

Recently, research into using Bidirectional Encoder Representations from Transformers (BERT) has been started. BERT has been developed by Google and is a transformer-based ML technique for NLP (232). It works by learning the contextual information in sentences by looking both backward and forwards (hence bidirectional) from words within a sentence. Examination of email headers and bodies using BERT shows that BERT performs significantly better than other methods (specifically THEMIS and THEMISb) (233), and similar results are shown when using BERT to examine URLs (234). Even more recently, XLNet has appeared on the scene (235,236), and several web articles state that it performs better than BERT (237–239). Although only a little research has been done so far, that which has been done shows that XLNet performs better than BERT in identifying phishing and spear-phishing emails (240,241).

Deep learning is sometimes combined with NLP (242–248).

A special mention should go to Sonowal and Kuppusamy as they have developed a system to help visually impaired people spot phishing attacks (249). Their process combines ML with typo squatting[33] and phoneme detection. Although their system works with a screen reader, it may be possible to adapt their method to help detect phishing emails

### 3.5.4. Strengths

AI and ML can be used to train the software to detect "Zero-day" phishing attacks with a high degree of accuracy.

### 3.5.5. Weaknesses

The primary perceived weakness is the time and human resources required to train an AI/ML system. Galinkin states that "*Machine learning is not magic! It requires a lot of maintenance and expertise to keep running*". He then suggests that "*If you can do something nearly as well without machine learning, you should probably not use it!*" (250).

In addition, ML should produce an objective, evidence-based decision-making process for detecting phishing attacks; however, (251) contends that "*machine learning (language processing) can acquire stereotyped biases from textual data reflecting everyday human culture*." They found that, even when trained on large sets of data, ML algorithms could still produce high numbers of false-positive results. They use the example of Aesop's fable "The Boy Who Cried Wolf"[34] to describe the likely outcome of this – end-users ignoring the warnings and, hence, becoming a victim of a phishing attack.

## 3.6. Proactive

This section reviews how proactive techniques can be used to restrict the effectiveness of phishing attacks.

The keywords and results of this search are detailed in Appendix B.

Proactive detection could be broadly defined as an attempt to detect phishing emails before they are read by the end-user. Using this definition, any technique that examines the contents of an email before the end-user reads it could be called

---

[33] https://www.kaspersky.com/resource-center/definitions/what-is-typosquatting

[34] http://read.gov/aesop/043.html

proactive.

However, for the purpose of this paper, proactive detection is being defined as the detection of new threats before they are delivered to the end-user. This would include identifying new, previously unknown phishing emails and websites – in effect, zero-day detection.

### 3.6.1. Proactive detection of phishing websites

(40) introduces this idea by stating that this method "*detects probable phishing URLs by generating different combinatorial URLs from existing authentic URLs and determining whether they exist and are involved in phishing-related activities on the web.*" (51,252,253) propose using existing URL blocklists to generate the new URLs to be examined, and (254) suggest creating URLs using brand identities (e.g., PayPal, Amazon, banks, etc.)

While this method would work, it appears to have several drawbacks:

- Each method requires heuristic rules to be defined to modify the selected URL. This could perhaps be mitigated with the use of ML techniques that learn how phishing URLs are altered to mimic genuine URLs.
- It is restricted to attempting to detect specified brands. Although it would be a trivial exercise to add new brands, it would require human intervention to do this.
- It is a rather brute force method and appears overly clumsy and time-consuming.

An alternative method is described in (255). It proposes examining every new domain as it is registered for phishing indicators. Many of these indicators are listed in Table II. Although this could work, several issues would have to be addressed:

- According to Verisign, 3.8 million new domain names were registered in the second quarter of 2021 (256), all of which would require checking. This equates to a little over 42,000 per day. Other sources suggest this number is on the low side – the WhoisDownload website[35] suggests that during September 2021, this was more than 120,000 per day, all of which would need to be identified and checked. This could potentially be mitigated through collective action and information sharing, for example, having several checking centres, each responsible for checking a subset of domain names.
- Some registrations are for holding purposes only, they may not have any hosting, or the host has a default or holding page only. Phishing websites may appear on these after some time, necessitating a recheck. This would imply that any domain name that could not be positively identified as legitimate would have to be regularly rechecked.
- Legitimate websites could be compromised, and these would never get checked.

However, this technique could identify many new phishing websites before they are used for malicious purposes.

### 3.6.2. Proactive detection of phishing emails

Several methods of real-time examination of emails have been proposed by (67,257–259), however as these all require the examination of emails before or during delivery to the end-user, they would be better classified as before-the-click detection rather than proactive.

The use of honeypots has been suggested by (257,260,261). A honeypot is defined as "*a decoy to lure cyber attackers and detect, deflect and study hacking attempts*" (262). There are several different types of honeypots, and the specific one of interest here is an email or spam trap which places "*a fake email address in a hidden location where only an automated address harvester will be able to find it*" (263). Any emails that are sent to that email address are automatically classified as spam or phishing. Other incoming emails can be compared against them to detect spam or phishing attacks.

Honeypots are used extensively as spam traps (264,262,265). (44) identified one paper referencing the use of a honeypot approach to proactively detect phishing attempts (266). Although this particular paper focuses on vishing, the technique could potentially be used to detect phishing attacks. (260) describes the use of honeypots within an e-banking system specifically designed to stop phishing attacks on client accounts, and (257) describes how honeypots can be used to detect and defend against malware attacks.

### 3.6.3. Strengths

Proactive detection can detect early use of phishing emails, potentially catching zero-day attacks.

Since all activity associated with honeypots is unauthorized, (261) asserts that they have four major strengths: no false positives (i.e. all activity is malicious), ability to capture new or unexpected behaviour (zero-day attacks), high-value datasets (all data is suitable for investigation), and the detection of encryption/protocol abuse.

### 3.6.4. Weaknesses

Proactive detection may be time and resource hungry.

According to (261), honeypots have three specific weaknesses: Limited view (can only report malicious activity directed against them), the possibility of the attacker identifying the honeypot and avoiding it, and the risks associated with deploying a honeypot on a live network (allowing potentially malicious emails access to a live system).

### 3.7. Other detection techniques

The keywords and results of this search are detailed in Appendix B.

### 3.7.1. Network round trip time

The literature review identified NRTT as an additional detection technique (30). According to (267), NRTT can be defined as "*the summation of the time a packet takes to travel from the server to the client and the time its acknowledgment takes to travel back from the client to the server.*" In other words, how long it takes to send a request and receive an

```
C:\WINDOWS\system32>ping google.co.uk

Pinging google.co.uk [142.250.179.227] with 32 bytes of data:
Reply from 142.250.179.227: bytes=32 time=18ms TTL=114
Reply from 142.250.179.227: bytes=32 time=17ms TTL=114
Reply from 142.250.179.227: bytes=32 time=17ms TTL=114
Reply from 142.250.179.227: bytes=32 time=21ms TTL=114

Ping statistics for 142.250.179.227:
    Packets: Sent = 4, Received = 4, Lost = 0 (0% loss),
Approximate round trip times in milli-seconds:
    Minimum = 17ms, Maximum = 22ms, Average = 19ms

C:\WINDOWS\system32>
```

*Fig. 2 Ping google.co.uk results*

acknowledgement. This process is easily demonstrated by using the "ping" command (Fig. 2)

For authentication purposes, NRTT can be used as a "fingerprint" of the genuine website for later verification.

(268) proposes using NRTT within cloud-based systems where there are multiple servers to determine which server will respond fastest for a user. In this environment, all the servers are pinged on first use, and the one with the lowest average NRTT is used for the rest of the session.

(30) proposes using NRTT to detect phishing websites. Their premise is that the NRTT of a phishing site will be different from that of the site that is being spoofed. They propose comparing the NRTT of the real site with the suspect site, and if they are different, then the suspect site is flagged as phishing.

The major problem with this process is what happens when authentication is required from another device, possibly in another location, as the NRTT could be significantly different due to any number of factors.

NRTT also shares the same problems as other techniques where the real and suspect sites are compared, that of determining what the real site is from examining the suspect site.

### 3.7.2. DMARC

Domain-based Message Authentication Reporting and Conformance (DMARC) is an email authentication, policy, and reporting protocol (269). It builds on the Sender Policy Framework (SPF) and Domain Keys Identified Message (DKIM) protocols. It relies on being implemented on the email sender's domain and being checked by the receiver of the email. If implemented correctly, it will confirm whether an email has been sent by the genuine user or whether the sender has been spoofed, in effect acting as a allowlist/blocklist for the email. According to (270–272), uptake of DMARC authentication is poor, with only 20% of enterprises having implemented it. Two reasons are given, the lack of understanding of the benefits and the perceived difficulties in implementing it. In the UK, the NCSC is on a mission to ensure that DMARC is implemented in all government departments. One of the primary drivers for this are the regular phishing campaigns being sent purporting to be form HMRC and this has resulted in a lack of trust by end-users when genuine emails are sent.

Where DMARC is correctly implemented, it accurately identifies genuine and phishing emails, and where the enforcement policy is also implemented and acted upon, phishing emails are quarantined or not delivered.

However, DMARC is not widely used, with perceived difficulties and lack of understanding of the benefits being the two most common reasons for it not being implemented (270–272).

### 3.8. Hybrid

The keywords and results of this search are detailed in Appendix B.

Hybrid is a combination of two or more of the above techniques, for example, proactive, AI/ML, heuristic, content and visual.

(39) states, "*In this technique different techniques are combined to detect if a website is fake or real.*"

The commonest combinations are heuristics and content and/or visual inspection while examining websites and web pages and blocklist/allowlist and heuristics while examining email content.

Its major strength is that multiple methods of detection can increase the probability of detecting phishing attacks.

However, it may also increase the number of false-positive results where legitimate emails and websites are marked as phishing.

## 4. Conclusion

This paper provides a systematic survey of anti-phishing detection techniques used in the real world and described in academic research. Most of the detection techniques focus on after-the-click detection. In the real world, before-the-click detection solutions are expensive and limited in use to large organizations which have the technical infrastructure available to implement them. This leaves SMEs and domestic users with no automated defences available for their use.

Part of before-the-click detection could include proactive detection, that is, the automated detection and classification of phishing websites and emails. This could improve before-the-click detection results, for example, by identifying zero-day attacks.

### 4.1. Before-the-click detection

This section will comment on the suitability of each of the after-the-click detection techniques for before-the-click detection of phishing emails. The focus is on examining the email before the end-user opens it.

It would be very easy to implement URL checking using blocklists, allowlists, and heuristics in before-the-click detection. The primary issue is that, like after-the-click detection, this is unlikely to discover zero-day attacks. The examination could be further complicated where the mal-actors use URL shorteners or redirects. The examination process will have to determine what the end destination of any URL is. This can be done and is just a coding exercise.

It will also be straightforward to implement heuristics checking for many of the email characteristics, including DMARC if it has been implemented (see Table III and Table IV Again, this would just be a coding exercise.

A majority of content and visual techniques are concerned with the examination of websites and hence could not be used to examine emails. Those that could potentially be used to examine emails include hashing and writing styles.

ML to create rules could be a useful technique. Research has primarily concentrated on using supervised learning algorithms, which require the input data to be tagged as phishing or not phishing. As these algorithms are resource and time-consuming, this technique may be best implemented to enhance, improve, or reinforce heuristic rules, perhaps as part of a reporting system. In such a system, the end-user can report an email as phishing (if it has not been so flagged) or as non-phishing (if it has been flagged as such). The email can then be passed through the ML system to help improve the detection rules.

Possibly the most promising technique would be the use of NLP. Here the language of the email would be examined, and NLP techniques used to determine whether the email is phishing or not. It currently appears that very little research has been done in this area.

## 4.2. Further research

This section provides some indication of potentially promising areas for future research and questions that could be answered, especially regarding before-the-click detection of phishing, spear-phishing, and whaling email attacks.

### 4.2.1. Blocklist/allowlist

Both blocklist and allowlist detection have a major flaw, that of the time taken to detect and classify URLs and update the appropriate list. There will always be a time gap where phishing URLs will remain undetected by this technique, hence allowing zero-day attacks. There is a need to reduce this time gap, perhaps by using proactive detection techniques.

Additionally, the URLs in the blocklist and allowlist need to be regularly rechecked. URLs in the blocklist can, and often do, disappear after a time, either because the phishing campaign has been completed, the mal-actors have moved the website to another host, or the website has been taken down by the authorities. URLs in the allowlist could be compromised and then used for phishing attacks by replacing the valid website with a phishing site.

Further investigation into the use of DNS and IP blocklists needs to be conducted to discover whether these are viable options for testing and to answer the questions "Can these blocklists be created and maintained? If so, how?" and "How effective are they?"

### 4.2.2. Heuristics

Several rules have been identified to examine URLs and websites. Some rules have been suggested for the examination of emails (Table III and Table IV). What additional rules could be effectively used to examine an email?

There are websites that can expand a shortened URL to its end destination; however, it appears that they can only deal with recognized URL shortener services. Can a URL expander be written that will expand any shortened URL and follow any redirection? Can this be incorporated into an automatic detection system?

Hashing has been proposed by (142,143) as a technique for comparing a suspect website with known phishing websites. Could this technique be used to compare suspect emails against known phishing emails? This will only work for identical emails, so what pre-processing of the email needs to be performed (for example, removing the greeting, removal of hidden/unprintable characters, etc.). A potential solution to this is perceptual hashing. Perceptual hashing produces "*a fingerprint of a multimedia file derived from various features from its content. Unlike cryptographic hash functions which rely on the avalanche effect of small changes in input leading to drastic changes in the output, perceptual hashes are "close" to one another if the features are similar.*" (273). It appears, from the authors' description, that perceptual hacking is designed to be used with image and video media. Further examination would be required to determine whether perceptual hashing could be used with emails.

### 4.2.3. Content/Visual

Do enough emails use a document object model, and if so, would DOM examination work?

Would document authorship work for larger organizations where end-users receive many emails from the same people? Could this technique be used to examine newsletter-type emails and identify fraudulent ones?

### 4.2.4. Artificial intelligence/machine learning

AI and ML have been popular areas for research, and many different AI/ML algorithms have been examined. However, this is an area where new algorithms are being regularly created. (274–278) list AI/ML algorithms that are new and popular. Most of these have already been investigated, but there appears to have been no research performed with respect to anti-phishing defences, for example, LightGBM, CatBoost, Linear Discriminant Analysis, Learning Vector Quantization, and Dimensionality Reduction Algorithms. (279) discusses the use of Bi-gram distance in name matching, which might also provide a fruitful area for future research.

Would any of the non-supervised ML algorithms (for example, Clustering, Association) provide satisfactory detection?

While most of the papers that researched the use of ML described the size of the datasets used for the training, none reported on how much time the training took. It is also difficult to compare the accuracy between different ML algorithms as the researchers used different datasets for training and testing. In most cases, the research appears to be unreproducible as the datasets are not made available, as highlighted by (280). This perhaps suggests another area of potential research; a comprehensive comparison of the accuracy, time taken, and resources needed to train each ML algorithm. Researchers could also benefit from having access to open-source datasets that they could use for training and testing their models. These datasets would have to be regularly maintained and updated.

ML produces rules for examining suspicious emails. One area of future research could compare ML rules with those used in the heuristic technique, checking which (if either) produce better results – highest true positive and true negative, lowest false positive and false negative. A second area could compare the time and resources required to develop, train, and use both techniques.

So far, little research has been done into the use of NLP to detect phishing attacks. Recently several new NLP models have been developed (281). These include several derivatives of BERT (RoBERTa, ALBERT, StructBERT, DeBERTa), OpenAI's GPT-2 (282) and GPT-3 (283), Efficiently Learning an Encoder that Classifies Token Replacements Accurately (ELECTRA) (284) and Google's Text-to-Text Transfer Transformer (T5) (285,286) and Switch Transformer (287,288). Google's Switch Transformer has been showing significant performance and training improvements over other models (289).

Could the use of NLP be used to detect phishing, spear-phishing, and whaling attacks?

### 4.2.5. Proactive

Proactive detection is the detection of new threats before they are delivered to the end-user. This would include identifying new, previously unknown phishing emails and websites – in effect, zero-day detection.

What would be the procedure for proactive detection?

How would the newly registered domain checking lifecycle work for proactive detection?

Phishing emails are not always the same. Could emails captured by honeypots be generalized?

Can content and visual examination of websites be incorporated into proactive detection?

Could NRTT be used to help proactively detect phishing websites by supporting content and visual inspection?

# Appendix A: Summary of interviews

**National Cyber Security Centre Suspicious Email Reporting Service**

Two interviews were conducted on 27th November 2020 and 1st December 2020 with three National Cyber Security Centre (NCSC) staff members to ascertain how the NCSC's Suspicious Email Reporting Service (SERS) worked. The interviews were conducted using Microsoft Teams (due to the Covid-19 pandemic, face-to-face interviews were not possible).

In April 2020, the NCSC launched SERS (290,291). Within two months of its launch, the NCSC had received one million reports of suspicious emails (292).

This service allows members of the public to report suspicious emails, which are then investigated, and the appropriate action taken.

The interviewees elaborated on the published process (290).

Emails are reported by forwarding them to the NCSC via their SERS email address, report@phishing.gov.uk. Personally identifiable information (PII) is then stripped from the emails (this includes the name and email address of the sender, along with any other information that could identify them, for example, signatures, links to social media accounts, etc.). The email is then sent on to the NCSC's takedown partner (at the time of writing, this is Netcraft), who performs the analysis of the email. This is primarily done by inspecting all the URL links in the email. If the email is found to be a phishing email, several processes are now performed. Firstly, the URLs pointing to the phishing site are added to a blocklist. This blocklist is made available to security software vendors (anti-virus, firewall, browser plugin developers, etc.) so that their software will block attempts to access the site. Secondly, if the email is spoofing a government website (gov.uk), the takedown provider attempts to get the hosting provider (wherever they are in the world) to take down the site; if this is impossible, the IP address of the site is blocked. Thirdly, for non-government spoofing websites that are hosted by a UK provider, the provider is contacted and requested to take the site down.

According to the people I interviewed, this request is usually acceded to and very often within a few minutes of the request being made.

Finally, data is aggregated, and analytics is provided to the police. These analytics include the brands being impersonated (e.g., PayPal, eBay, Amazon, high street banks, etc.).

No feedback is provided to the end-user as to what action has been taken.

Unfortunately, taking down phishing websites is like "playing whack-a-mole"; as soon as one site is removed, it reappears elsewhere with a new IP address and domain name.

**Strengths of the service**

The service is government-backed, and hence UK based hosting services can be compelled to acquiesce to requests from the NCSC and their takedown partners.

According to the NCSC Annual Review 2020, from the start of the service in April 2020 to the report publication date, the takedown service successfully took down 99.6% (166,710) of identified phishing URLs; 65.3% of these were removed within 24 hours of them being determined as malicious. 42,574 phishing URLs were associated with UK Government themed attacks (293)

**Weaknesses of the service**

Due to the large number of phishing emails purporting to be from Her Majesty's Revenue and Customs (HMRC), end-users are now highly distrustful of any emails coming from HMRC and are either reporting or ignoring legitimate emails from HMRC. The end-user is not informed if the email reported to the NCSC turns out to be legitimate.

The service relies on end-users to spot potential phishing emails and report them. Unfortunately, humans are notoriously bad at spotting phishing emails.

The service does not integrate into the end-user's email system; instead, it relies on third-party security software (e.g., anti-virus, firewall, browser plugins, etc.) to block links when they are clicked on. Because of this, it will not stop spear-phishing or whaling attacks where the end-user receives an email urging them to take manual action (for example, transferring funds to a scammer's bank account).

The blocklist only lists known malicious sites. There is no "allow" list where sites that are verified good can be listed (for example, emails from HMRC). Also, there is a time delay between a malicious site being reported and it being added to the blocklist.

Only 10-15% of emails reported are phishing emails.

**Anti-Phishing Working Group**

Interview with Foy Shiva on 4[th] January 2021 to ascertain what the Anti Phishing Working Group (APWG) does and how they may be able to help other researchers. The interview was conducted using Zoom as Foy is in the USA.

The APWG's strapline is "Unifying the global response to cybercrime through data exchange, research and promoting public awareness." They have two areas of cybercrime that they specialize in, cryptocurrency and phishing. It is the phishing that we discussed.

The APWG works with member organizations[36] to gather and disseminate information about phishing emails and websites. Each reported suspicious link is assigned a confidence label (100% confirmed, 90% almost sure, probably reported through ML algorithms, 50% not yet vetted), along with the brand that is being spoofed, date discovered, hosting IP address, and other potentially useful information.

Members have real-time access to this information for use in their own anti-phishing solutions. No attempt is made by the APWG to take down the reported phishing websites, although a few members do attempt this.

The APWG can provide data dumps of selected sets of information for researchers to use. The APWG also has a research partnership program where university researchers can access and contribute to the phishing database[37].

**Strengths of the service**

The APWG works with several high-profile partners to collect, examine, and disseminate information about phishing websites and emails. This information is made available, in real-time, to their partners and researchers.

**Weaknesses of the service**

The APWG has no authority to take down malicious sites. Commercial partners pay for the service (up to $15,000/year).

# Appendix B: Literature search keywords and results

### Step 1: Keyword search

The keywords for the initial search were "phishing" (in the title, abstract, or keywords) and "literature survey" OR "state of the art" OR "SoK" OR "Systematization of Knowledge" (in all available search fields).

This search returned 349 records, of which 173 were removed as duplicates. Of the remaining 176 records, 162 were removed as they were not literature reviews, surveys, systematization of knowledge (SoK), or state-of-the-art publications. Of the final 14 publications, two were abstracts only. Requests were made to the authors for copies of the full text using the ResearchGate request service. At the time of writing, none of the authors had responded. The remaining 12 publications were considered suitable for review (30,34–36,38–43,101,294).

After reviewing the remaining 12 publications, one was discarded (294) as it provided no useful information.

### Step 2: Backward search

After the initial keyword search and screening of the papers identified in step 1, a backward search was performed looking for literature reviews referenced by those already identified. This search found several of the papers already identified but led to no new papers to investigate.

### Step 3: Forward search

Following the backward search, a forward search for citations of the publications identified in the initial search was performed. The databases searched included the ACM Digital Library, Google Scholar, IEEE Xplore[38], JSTOR[39], ScienceDirect, and Scopus (as suggested by Emory Library Ask a Librarian[40]) and the De Montfort University library.

The citation search produced over 800 papers, most of which were excluded as they were not general literature reviews. Note that this list did include literature reviews for specific detection techniques and will be revisited later.

23 papers were identified for further review – seven were only available for purchase or through a subscription that DMU does not have. However, one of these papers, "Web-Based Classification for Safer Browsing" (295), has made the abstract and references available online[41] , and there are some potentially interesting references to different detection techniques.

17 papers were examined, eight were excluded as not adding anything to this stage of the literature review, although they may be useful later for specific detection techniques.

The final nine (37,44–47,97,99,203,295) added to the descriptions and knowledge of the detection methods identified in the original search, as well as introducing additional techniques and methods to explore.

A further forward search of papers citing these nine produced no new useful papers.

### Step 4: Keyword search for each of the identified detection techniques

Further searches of the De Montfort University, IEEE, ACM, ProQuest, Wiley, and Science Direct libraries, along with Google Scholar, were made for keywords associated with each of the identified detection techniques. The details of each search are included in the section about that detection technique. The Scopus library search was dropped as all the search results were being returned from other libraries, and Scopus was only returning abstracts of the papers.

**Blocklist/allowlist detection technique search.** The keywords used for the literature search for this detection technique were "phishing" (in the title) and "blacklist" OR "whitelist" (in the title, abstract, and keywords fields). In addition, a search was done for papers that compared the effectiveness of phishing blocklists.

This search returned 606 records, of which 535 were removed as they were primarily proposing alternatives to the blocklist/allowlist technique. Of the remaining 71, 12 were duplicates. Of those remaining, 23 examined the use of blocklists and allowlists (39,50,52,53,55,57,59,66,166,252,296–308), 11 examined the use of blocklists or allowlists as part of a hybrid approach (51,63,66,77,297,299,309–313) and 25 identified problems with the blocklist/allowlist detection technique (56,60,117,254,300,308,314–332). An additional paper from the initial literature review search was also included for examination (40).

**Heuristic detection technique search.** The keywords for this detection technique were "phishing" (in the title) and "heuristics" OR "rules" OR "stateless" OR "search engine" OR "url" OR "html" OR "domain name" OR "string matching" OR "lexical" OR (("email" OR "e-mail") AND ("header" OR "body")) (in the title, abstract, and keywords fields). An additional search for "phishing" (in the title) and "Unicode" OR "homograph" (in the title, abstract, and keywords fields) was performed as it was referenced in the book "Phishing and Countermeasures: Understanding the Increasing Problem of Electronic Identity Theft (333).

This search returned 1553 records, of which 1427 were rejected as they were not specifically dealing with the heuristic detection technique. Of the remaining 126 records, 67 were duplicates. Of those remaining, 23 records addressed the examination of email content (including URL examination) (59,69,79–86,88–96,98,100,102,103), 16 records addressed the examination of web pages or website content (80,96,105–118), 16 records examined the use of heuristics as part of a hybrid approach (63–68,70–78,111), six records addressed the issue of Unicode or homograph attacks (102,123–126,334) and two records addressed the comparison of URLs with real URLs (119,120).

**Content detection technique search.** The keywords for this detection technique were "phishing" (in the title) and "content" OR "td-idf" OR "document object model" OR "dom" (in the title, abstract, and keywords fields).

This search returned 629 records, of which 585 were rejected as they were not specifically dealing with detecting content using non-heuristic methods. Of the remaining 43 records, 17 were duplicates. 26 records addressed the examination of content using non-heuristic methods.

15 records addressed the examination of content only (115,130–133,136,138–143,148,149,335).

11 records addressed examining content as part of a hybrid approach using ML (72,133,150,336–339), blocklist/allowlist (313), heuristics (313,340), search engines (131) and the semantic web (341).

**Visual detection technique search.** The keywords for this detection technique were "phishing" (in the title) and "visual" AND ("content" OR "similarity") (in the title, abstract, and keywords fields).

This search returned 158 records, of which 82 were rejected as they were not specifically dealing with visual detection methods. Of the remaining 76 records, 40 were duplicates. 36 records addressed the visual examination of websites.

25 records addressed visual examination only (137,150,153–155,158,160,167,168,342–357).

10 records addressed visual examination as part of a hybrid approach using ML (164,165), blocklist/allowlist (161,162), heuristics (152,162,163), and content (150–152).

One record highlighted the disadvantages of the visual examination of websites (166).

**Artificial intelligence/machine learning detection technique search.** The keywords for this detection technique were "phishing" (in the title) and "machine learning" OR "deep learning" OR "artificial intelligence" OR "neural networks" OR "naive bayes algorithm" OR "fuzzy logic" OR "natural language processing" OR "semantic network" OR "character N-grams" OR "data mining" OR "support vector machine" OR "adaboost" (in the abstract or subject terms/keywords).

At the suggestion of Dr Iryna Yevseyeva (first supervisor for this research), an additional search was done for Bidirectional Encoder Representations from Transformers (BERT). This search used the keywords "phishing" (in the title) and "bidirectional encoder representations from transformers" OR "BERT" (in the abstract or subject terms/keywords).

This search returned 1559 records of which 1419 were rejected as they were not specifically dealing with the AI/ML detection technique or were duplicates. Of the remaining 140 records, 70 records addressed the use of ML, of which 61 used feature extraction (72,74,76–78,114,118,133,150,180,249,315,317,319,320,326–328,330,338,358–398) and 9 other methods of ML (255,306,399–405). 24 records examined the use of deep leaning and or neural networks (242–248,406–422). 25 records compared the results of using different ML algorithms, one of which has since been retracted(74,164,179–200). 20 records examined the use of NLP, semantic networks, phonetics or BERT (187,230,231,233,234,240,249,311,327,329,386,423–431) and one highlights problems and issues with the use of AI/ML (251).

Google Scholar was not used in this search as it returned more than 10,000 results, and an examination of the first few pages showed that they were all duplicates of results already retrieved.

The other libraries returned only duplicates already found in the DMU library search.

**Proactive detection technique search.** The keywords for this detection technique were "phishing" (in the title) and "spam trap" OR "domain name registration" OR "proactive detection" OR "predictive blacklist" OR "proactive blacklist" (in the title, abstract, and keywords fields).

This search returned 49 records, of which 27 were rejected as they were not specifically dealing with the heuristic detection technique. Of the remaining 22 records, 12 were duplicates. Of those remaining four proposed creating potentially suspicious URLs and checking for their existence (51,252–254), two proposed using honeypots (257,260), one examined newly registered domains (255), and four performed real-time examination of emails (67,257–259).

**Other detection techniques search.** Two separate searches were performed.

The first search was performed to identify where NRTT is currently referenced. The keywords for this search were "clas"

OR "communications latency authentication scheme" OR "network communications latency" OR "network latency profiling," OR "network round trip time" OR "round trip communications latency" (in the title, abstract, or keywords).

The search returned 627 records, of which 614 were rejected as they were not specifically dealing with the use of NRTT. Of the remaining 13 records, 8 were duplicates. Of those remaining four showed how NRTT could be used in online authentication (267,432–434), and one with cloud server selection (268).

The second search was performed to determine where NRTT is being specifically referenced in connection to phishing. The keywords for this search were "phishing" (in the title, abstract, or keywords) and "clas" OR "communications latency authentication scheme" OR "network communications latency" OR "network latency profiling" OR "network round trip time" OR "round trip communications latency" (in the title, abstract, keywords or text).

The search returned one result, the already identified (30).

## Acknowledgement

Trevor Wood would like to thank his company, Network Midlands Ltd, for providing time and funding towards producing this paper, the three staff members at the UK's National Cyber Security Centre and Foy Shiva of the APWG for their interviews and insights, and his prayer support group.

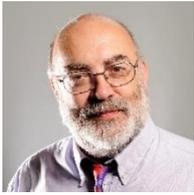

Trevor Wood received his first MSc in Information Systems from Kingston University, London, UK in 1999 and his second MSc in Digital Forensics and Cybersecurity from De Montfort University, Leicester, UK in 2019. He is currently studying for a PhD in Cybersecurity at De Montfort University, Leicester, UK where he is researching the use of Natural Language Processing to detect advanced phishing emails. He is a director at Network Midlands Ltd, Leicester, UK, an internet and cybersecurity consultancy. He has over 45 years of experience in IT, working for the National Physical Laboratory, the BBC, British Gas Research, BP and consulting with many SME and SOHO businesses. He is a co-author of "How Location-Aware Access Control Affects User Privacy and Security in Cloud Computing Systems". He is especially interested in all aspects of social engineering.

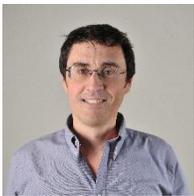

Vitor Basto Fernandes graduated in information systems in 1995 (including an internship at Ascom Tech AG - Switzerland), post-graduated in distributed systems in 1997 and got his PhD on multimedia transport protocols in 2006, all from the University of Minho (Portugal), where he has also been teaching assistant. He worked at Integral Vision Inc (UK) as a software quality engineer in 1996 and was co-founder of PSI-Information Systems Lda (Portugal) in 1997, where he was project manager in B2B e-commerce web-based software development. He was invited assistant professor at the University of Tras-os-Montes e Alto Douro (Portugal) in 2007 and 2008. In 2008 he joined the Informatics Engineering Department of Polytechnic Institute of Leiria (Portugal) as an adjunct professor, where he has been coordinator professor between 2014 and 2016. He coordinated the MSc Program in Mobile Computing from 2010 to 2012, was head of the Research Center in Computer Science and Communications at Polytechnic Institute of Leiria between 2012 and 2016, and researcher in several international projects in the areas of information systems integration, anti-spam filtering and multiobjective optimization. Vitor joined the Department of Information Science and Technology at the University Institute of Lisbon (Portugal) in 2016, where he is currently an assistant professor with habilitation.

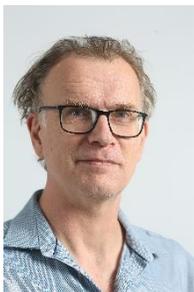

Eerke Boiten graduated ("ir.") in Computer Science at the University of Twente in 1988 and obtained a PhD in Computer Science at the (then) Catholic University of Nijmegen in 1992. His PhD research in transformational programming formed the start of a long research career in the use of mathematics for the construction of correct programs or "formal methods". He moved to the University of Kent in 1995 and, while there, made notable contributions in the areas of viewpoint specification, refinement, and the combination of state-based and behavioural specification. He co-authored two monographs on refinement and chaired the Refinement Workshop for 15 years. His research shifted in the mid-2000s towards the application of formal methods to security and subsequently to broader cyber security and privacy. Eerke took on the leadership of the Security research group at Kent and developed it into an interdisciplinary centre that received GCHQ/EPSRC accreditation as an Academic Centre of Excellence in Cyber Security (ACE-CSR). In 2017 he moved to De Montfort University to take a chair in Cyber Security and lead its Cyber Technology Institute, for which he led the successful ACE-CSR accreditation in 2019. In that same year, he became Head of the School of Computer Science and Informatics. He contributes to public debate on privacy and security on a regular basis, including with around 40 yearly appearances in press and media in recent years.

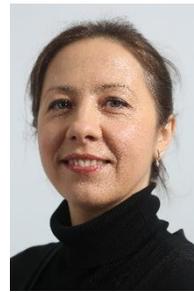

Iryna Yevseyeva is an Associate Professor in Computer Science and Deputy Subject Group Leader for Cyber Security at the School of Computer Science and Informatics at the Faculty of Computing, Engineering and Media at the De Montfort University, Leicester, UK. She is a member of DMU's Cyber Technology Institute. She obtained her PhD from the University of Jyvaskyla, Finland, in 2007 on multicriteria decision aiding approaches for classification and her MSc in Mobile Computing from the same university in 2001. Her research interests are in computational intelligence, (multiobjective / multicriteria) decision making and aiding techniques, and their application in various domains including cyber security.